\documentclass[3p,times,procedia]{elsarticle}
\usepackage{nupha_ecrc}

\volume{00}

\firstpage{1}

\journalname{Nuclear Physics A}

\runauth{}

\jid{nupha}

\jnltitlelogo{Nuclear Physics A}

\usepackage{amssymb}
\usepackage{lineno}

\biboptions{sort&compress}

\usepackage[figuresright]{rotating}

\begin{document}

\begin{frontmatter}



%% Instructions from Editor: Please use the following \dochead only in the preprint version (e-print arXiv etc.); 
%% use empty \dochead{} when submitting to Nuclear Physics A!
\dochead{XXVIIIth International Conference on Ultrarelativistic Nucleus-Nucleus Collisions\\ (Quark Matter 2019)}
%% e.g. \dochead{17th International Conference on Dynamical Processes in Excited States of Solids}

\title{Measurement of the Sixth-Order Cumulant of
Net-Proton Distributions in Au+Au Collisions
from the STAR Experiment}


\author{ Toshihiro Nonaka for the STAR Collaboration }

\address{ Central China Normal University, Wuhan 430079, China \\ University of Tsukuba, Tsukuba, Ibaraki 305, Japan }

\begin{abstract}
Higher-order cumulants of the net-baryon multiplicity distributions are predicted to be 
sensitive to the properties of the nuclear matter created in high-energy nuclear collisions.
In this talk, we present the collision centrality and acceptance (rapidity and transverse momentum) dependence of the ratio of the 6$^{{\rm th}}$- 
to the 2$^{\rm nd}$-order cumulant ratio ($C_{6}/C_{2}$) 
of net-proton in Au+Au collisions at $\sqrt{s_{\rm NN}}=54.4$ and 200~GeV measured by the 
STAR detector at RHIC.
The new results are compared to hadron transport model and lattice QCD calculations. 
\end{abstract}


\end{frontmatter}

%\linenumbers

% \linenumbers

\section{Introduction}
One of the main goals of heavy-ion collision experiments is to study 
the phase diagram of Quantum Chromodynamics (QCD).
Higher-order cumulants of conserved charges are predicted to be sensitive to the 
QCD phase structure, especially the possible critical end point.
The STAR experiment has investigated the QCD phase structure at finite $\mu_{\rm B}$ 
by measuring the cumulants up to the 4$^{\rm th}$ order 
($C_{n}, n<4$) and their ratios in the STAR experiment at RHIC~\cite{Aggarwal:2010wy,net_proton,net_charge,net_kaon}.
Recently, the non-monotonic beam-energy dependence is observed for $C_{4}/C_{2}$ of net-proton 
multiplicity distributions~\cite{Adam:2020unf}, which could be an experimental signature of the critical point.
At $\mu_{\rm B}=0$~MeV, on the other hand, the lattice QCD calculation predicts a smooth crossover ~\cite{Aoki:2006we}, 
although there is no direct experimental evidence of a crossover. 
Theoretically, the 6$^{\rm th}$- to 2$^{\rm nd}$-order cumulant ratio, $C_{6}/C_{2}$, of the net-charge and net-baryon 
multiplicity distributions are predicted to be negative if the freeze-out temperature is close to the 
pseudocritical temperature~\cite{Friman}. 

In this proceedings, we present the centrality dependence of $C_{6}/C_{2}$ of net-proton multiplicity 
distributions at $\sqrt{s_{\rm NN}}=54.4$ and 200~GeV. The results are compared with the hadron transport 
model calculations and lattice QCD calculations.
The acceptance dependence of $C_{6}/C_{2}$ is also discussed.

\section{Analysis methods}
The data are collected by STAR using the Time Projection Chamber (TPC) and Time-Of-Flight (TOF) detector. 
Collision events occurred within 30~cm in the beam direction and 
2~cm in the radial direction are selected.
Protons are identified by TPC and TOF at rapidity window $-0.5<y<0.5$ 
and transverse momentum window $0.4<p_{\mathrm T}\;({\rm GeV/c})<2.0$. 
The average PID efficiency is about 50\% with a weak centrality dependence~\cite{Adam:2020unf}.
Centrality is determined in $|\eta|<1.0$ excluding protons and antiprotons 
to gain the maximum centrality resolution with reducing the autocorrelation effects~\cite{Luo:2013bmi,Luo:2017faz}. 
The centrality bin width averaging (CBWC) is done in order to suppress the effect of 
the initial volume fluctuations~\cite{Sugiura:2019toh}.
Cumulants are calculated for each multiplicity bin, which are averaged in 10\% step of the centrality bin. 
In order to reduce the statistical uncertainties in central collisions, 
results in 0-10, 10-20, 20-30 and 30-40\% centrality intervals are averaged using the inverse 
of the error squared as a weight. 

Efficiency and acceptance corrections are done assuming that the detector 
responses do not depend strongly on the event multiplicity~\cite{eff_kitazawa,eff_koch,eff_xiaofeng,Nonaka:2017kko,Luo:2018ofd}. 
Within the aforementioned acceptance, the efficiency for antiprotons is found to be lower than 
that of protons by $\sim$5\% independent of centrality. 
Unlike the case for particle yield analysis, the efficiency correction of the higher-order cumulants 
is expressed by the convolutions of lower-order cumulants, which means that 
the correction is affected not only by the value of 
the single-particle efficiency but also strongly affected by the shape of the distributions. 
%For example, the correction for the 4$^{\rm th}$ order net-proton cumulants 
%include both the mean and sigma of the distributions~\cite{eff_koch}.

\section{Centrality dependence}
\begin{figure}[htbp]
  \begin{center}
   \includegraphics[width=10.0cm]{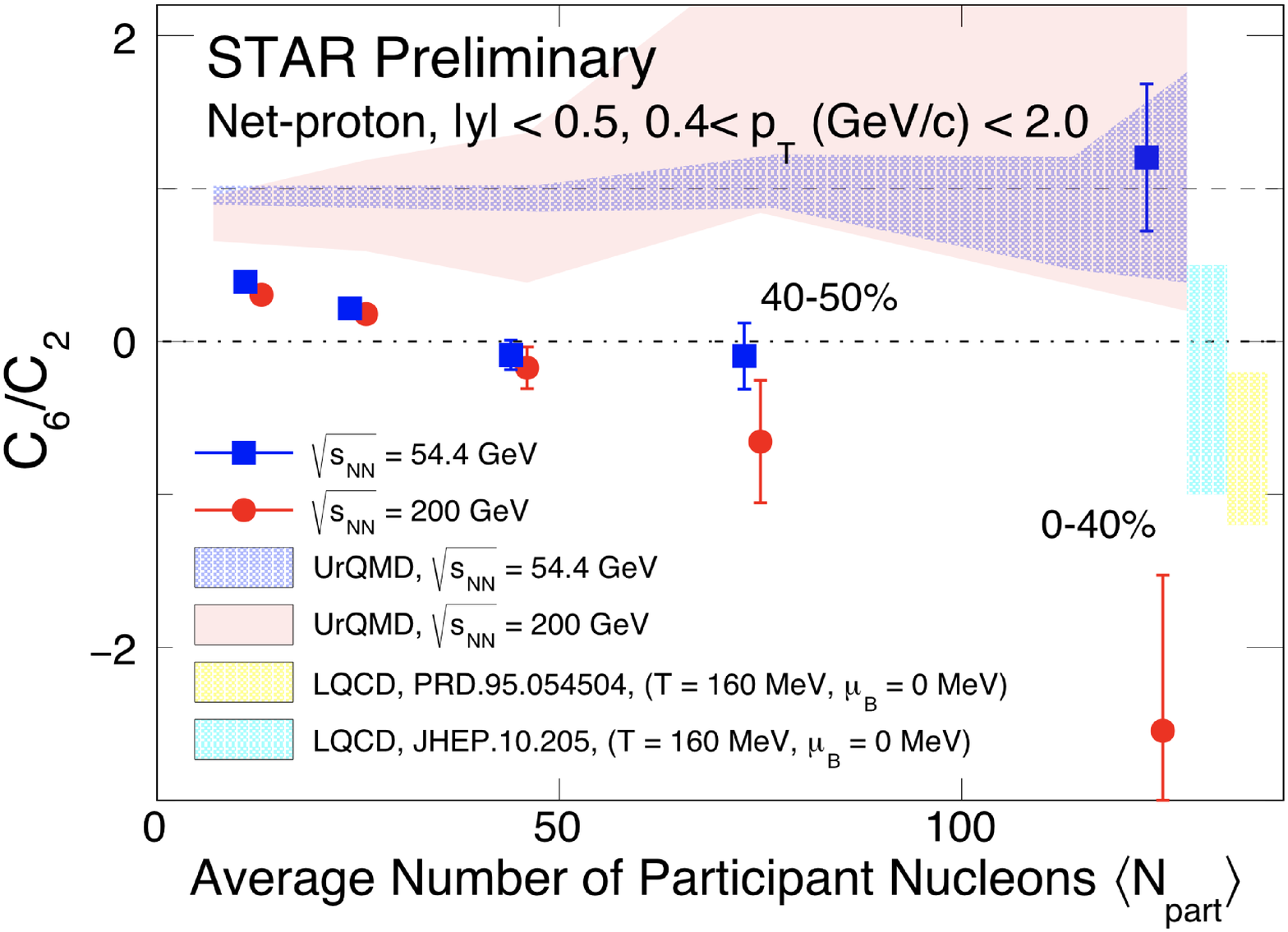}
  \end{center}
  \vspace{-0.3cm}
  \caption{
	  Centrality dependence of $C_{6}/C_{2}$ of net-proton multiplicity distributions 
	  in Au+Au collisions at $\sqrt{s_{\rm NN}}=54.4$ and 200~GeV. The most central bin represents 0-40\% centrality.
	  The hadron transport model (UrQMD) calculations are shown in blue and red bands for two beam energies.
	  The cyan and yellow bands show LQCD calculations taking $T=160$~MeV.
	  }
  \label{fig:centdep}
\end{figure}

Figure~\ref{fig:centdep} shows the centrality dependence of $C_{6}/C_{2}$ of net-proton multiplicity 
distributions in Au+Au collisions at $\sqrt{s_{\rm NN}}=54.4$ and 200~GeV. 
It is found that the results from two beam energies are consistent in peripheral 
collisions, while a clear separation is observed in 0-40\% central collisions (the most central bin). The negative sign is 
observed for $\sqrt{s_{\rm NN}}=200$~GeV. The $C_{6}/C_{2}$ value at $\sqrt{s_{\rm NN}}=54.4$~GeV 
shows positive sign and consistent with the Skellam baseline (=1). 
The hadron transport model (UrQMD) calculations, which does not incorporate the phase transition, 
show positive sign of $C_{6}/C_{2}$ of net-proton multiplicity distributions 
for all the centralities at $\sqrt{s_{\rm NN}}=54.4$ and 200~GeV. 
The lattice QCD (LQCD) calculations from two groups taking $T=160$~MeV ($\mu_{\rm B}=0$~MeV) 
also show negative values~\cite{Bazavov:2017dus,Borsanyi:2018grb,Bazavov:2020bjn}.
When we compare LQCD results with our measurements in 0-40\% central collisions at $\sqrt{s_{\rm NN}}=200$~GeV, 
we find that those results are consistent within large uncertainties.

\section{Acceptance dependence}
\begin{figure}[htbp]
  \begin{center}
   \includegraphics[width=11.0cm]{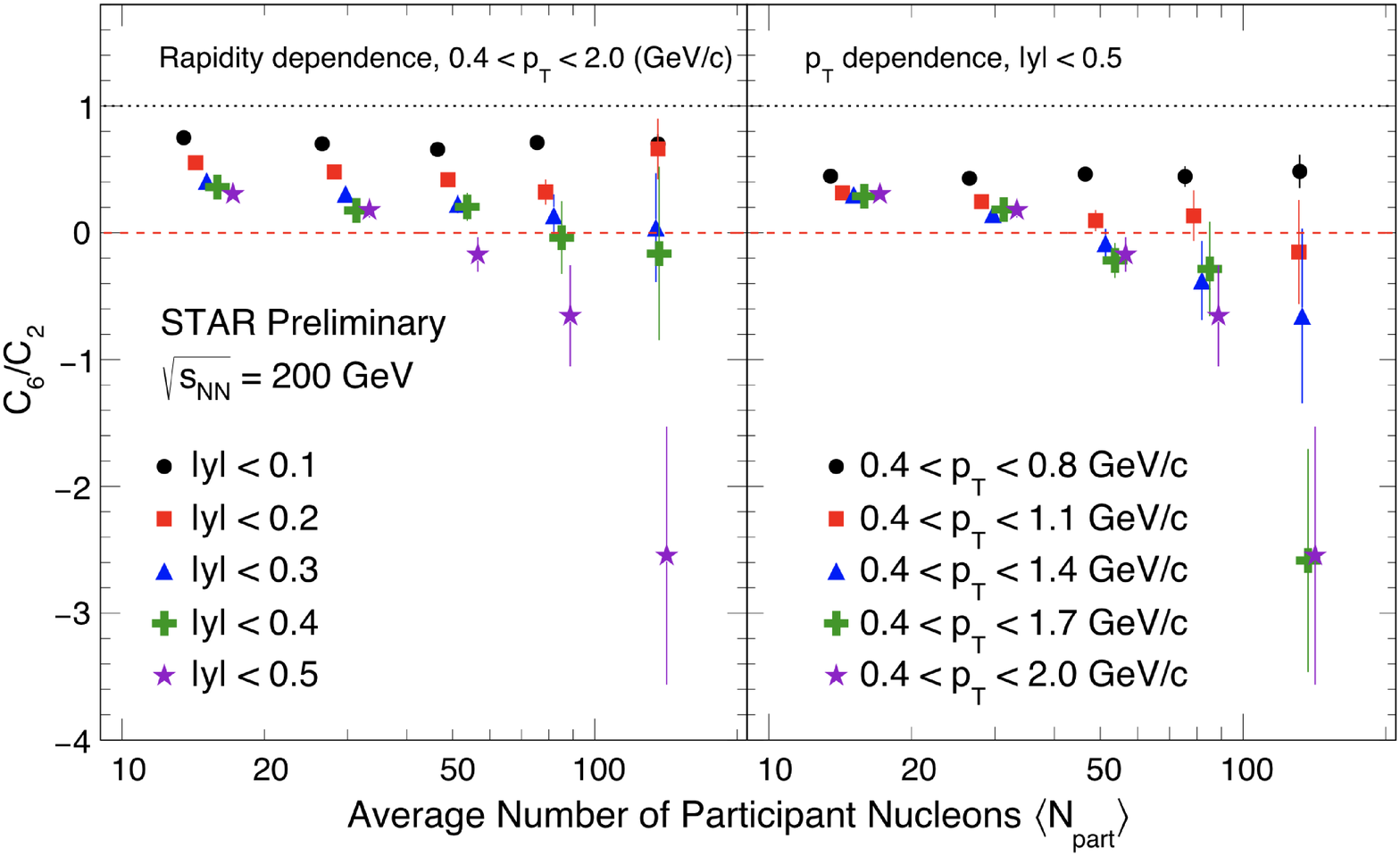}
  \end{center}
  \vspace{-0.3cm}
  \caption{
	  Rapidity ($y$) and transverse momentum ($p_{\mathrm T}$) acceptance dependence of $C_{6}/C_{2}$ 
	  of net-proton multiplicity distributions 
	  Au+Au collisions $\sqrt{s_{\rm NN}}=200$~GeV in 0-40\% centrality.
	  }
  \label{fig:accdep}
\end{figure}

When we compare the experimental results with LQCD calculations, we need to consider the effects of the experimental 
acceptance. With the narrow acceptance the net-baryon multiplicity distributions should reach the Skellam distribution, 
which exhibits $C_{6}/C_{2}=1$. By increasing the acceptance, the physics signals 
increase, which may lead to deviations from unity. 
The $C_{6}/C_{2}$ is expected to reach $0$ with the full $4\pi$ acceptance, 
where the net-baryon number is no longer fluctuate. 
Figure~\ref{fig:accdep} shows the acceptance dependence of $C_{6}/C_{2}$ of net-proton multiplicity 
distributions in Au+Au collisions $\sqrt{s_{\rm NN}}=200$~GeV for each centrality bin, 
where the rapidity window ($|y|<\alpha$, $\alpha=0.1$, 0.2, 0.3, 0.4 and 0.5) and $p_{\mathrm T}$ acceptance ($0.4<p_{\mathrm T}<\beta$, $\beta=0.8$, 1.1, 1.4, 1.7 and 2.0~GeV/c) 
is enlarged from left to right along the x-axis.
For both the rapidity and $p_{\mathrm T}$ acceptance dependence, the $C_{6}/C_{2}$ values 
are close to the Skellam baseline 
when the acceptance is small, but the values decrease for larger acceptance. 
The negative sign is observed for large acceptance 
in central collisions. It is found that the values are saturated for $p_{\mathrm T}$ acceptance dependence at $\beta>1.4$. 
This is because the mean $p_{\mathrm T}$ of protons is at around $p_{\mathrm T}\approx1.1$~GeV/c, hence 
the number of protons is saturated around $\beta>1.4$~GeV/c.

\section{Summary}
We report the first measurements of the $C_{6}/C_{2}$ of net-protons from 54.4 and 200~GeV Au+Au 
collisions at RHIC. The results from peripheral collisions at both energies are consistent with each other. 
The $C_{6}/C_{2}$ measured in 0-40\% Au+Au collisions at $\sqrt{s_{{\rm NN}}}=200$~GeV 
is found to be negative and consistent with the lattice QCD prediction. 
This consistency implies that the nuclear matter created at 200~GeV collisions might undergo a smooth crossover 
transition from the QGP to hadronic matter.
On the other hand, the result from 54.4~GeV shows a positive value. 
%Both are at 2 sigma effect. 
The negative and positive signs at 200 and 54.4~GeV in 0-40\% Au+Au collisions are at 2 sigma effect.
High statistics data are needed in order to understand the phase structure at vanishing baryonic density.

\section{Acknowledgement}
This work was supported by the National Key Research and Development Program of China (2018YFE0205201), the National Natural Science Foundation of China (No.11828501, 11575069, 11890711, 11861131009 and 11950410505), and China Postdoctoral 
Science Foundation funded project 2018M642878.



%% References


\bibliographystyle{elsarticle-num}
\bibliography{main}






% \end{thebibliography}

\end{document}